\documentclass[%
 reprint,
runinaddress,
longbibliography,
nofootinbib,
 amsmath,amssymb,
 aps,
]{revtex4-2}

\usepackage{graphicx}
\usepackage{dcolumn}
\usepackage{bm}
\usepackage{physics}
\usepackage[colorlinks=true, linkcolor=blue, citecolor=cyan]{hyperref}

\begin{document}

\preprint{APS/123-QED}

\title{Probing Ensemble Properties of Vortex-avalanche Pulsar Glitches with a Stochastic Gravitational-Wave Background Search}

\author{Federico De Lillo}
\email{federico.delillo@uclouvain.be}
\affiliation{Centre for Cosmology, Particle Physics and Phenomenology (CP3),\\
Universit\'e catholique de Louvain, Louvain-la-Neuve, B-1348, Belgium}
\author{Jishnu Suresh}%
\email{jishnu.suresh@uclouvain.be}
\affiliation{Centre for Cosmology, Particle Physics and Phenomenology (CP3),\\
Universit\'e catholique de Louvain, Louvain-la-Neuve, B-1348, Belgium}
\author{Antoine Depasse}%
\email{antoine.depasse@uclouvain.be}
\affiliation{Centre for Cosmology, Particle Physics and Phenomenology (CP3),\\
Universit\'e catholique de Louvain, Louvain-la-Neuve, B-1348, Belgium}
\author{Magdalena Sieniawska}%
\email{magdalena.sieniawska@uclouvain.be}
\affiliation{Centre for Cosmology, Particle Physics and Phenomenology (CP3),\\
Universit\'e catholique de Louvain, Louvain-la-Neuve, B-1348, Belgium}
\author{Andrew L. Miller}%
\email{andrew.miller@uclouvain.be}
\affiliation{Centre for Cosmology, Particle Physics and Phenomenology (CP3),\\
Universit\'e catholique de Louvain, Louvain-la-Neuve, B-1348, Belgium}
\author{Giacomo Bruno}%
\email{giacomo.bruno@uclouvain.be}
\affiliation{Centre for Cosmology, Particle Physics and Phenomenology (CP3),\\
Universit\'e catholique de Louvain, Louvain-la-Neuve, B-1348, Belgium}

\date{\today}

\begin{abstract}
A stochastic gravitational-wave background (SGWB) is expected to be produced by the superposition of individually undetectable, unresolved gravitational-wave (GW) signals from cosmological and astrophysical sources. Such a signal can be searched with dedicated techniques using the data acquired by a network of ground-based GW detectors. In this work, we consider the astrophysical SGWB resulting from pulsar glitches, which are sudden increases in the rotational pulsar frequency, within our Galaxy. More specifically, we assume glitches to be associated with quantized, superfluid, vortex-avalanches in the pulsars, and we model the SGWB from the superposition of GW bursts emitted during the glitching phase.
We perform a cross-correlation search for this SGWB-like signal employing the data from the first three observation runs of Advanced LIGO and Virgo. Not having found any evidence for a SGWB signal, we set upper limits on the dimensionless energy density parameter $\Omega_{\mathrm{gw}}(f)$ for two different power-law SGWBs, corresponding to two different glitch regimes. We obtain $\Omega_{\mathrm{gw}}(f)\leq 7.5 \times 10^{-10}$ at 25 Hz for a spectral index 5/2, and $\Omega_{\mathrm{gw}}(f)\leq 5.7 \times 10^{-17}$ at 25 Hz for a spectral index 17/2.
We then use these results to set constraints on the average glitch duration and the average radial motion of the vortices during the glitches for the population of the glitching Galactic pulsars, as a function of the Galactic glitch rate.

\end{abstract}
\maketitle

\section{\label{sec:introduction}Introduction}
The third observing run (O3) of the LIGO-Virgo-KAGRA~\cite{2015CQGra..32g4001L,2015CQGra..32b4001A,KAGRA:2020tym} collaboration finished by cataloging several tens of Gravitational Waves (GWs)~\cite{LIGOScientific:2021djp} originating from the compact binary coalescence (CBC) of black holes and/or neutron stars (NS). However, CBCs are only one class of GW sources among a broader range of possibilities. 
One of the interesting source categories yet to be detected is the stochastic gravitational-wave background (SGWB). The SGWB is considered to be a persistent signal resulting from the incoherent superposition of GWs from a large number of sources with cosmological and astrophysical origins. From the astrophysical perspective, there could be several phenomena contributing to the SGWB~\cite{Regimbau:2011rp,Regimbau:2022mdu}. These include the superposition of continuous gravitational waves from NSs~\cite{Dhurandhar:2011qe,lambda_statistic,DeLillo:2022blw,Agarwal:2022lvk}, magnetars~\cite{cutler,marassi_magnetar,WuMag,bonazzola,BenjOwn}, core-collapse to supernovae bursts~\cite{Crocker1,Crocker2,Finkel,Buonanno_SN,Marassi_Pop23,Coward_SN}, and the superposition of the unresolved astrophysical CBC events~\cite{regfrei,zhu_cbc,marassi_cbc,rosado,regman,wu_cbc}. Even though these astrophysical phenomena can be classified as sources of weak GW signals, their collective and incoherent signals will form a SGWB, and we may be able to observe them with the network of ground-based GW detectors.

From the high precision tracking of the pulsar spins (for a review, see \cite{Lorimer:2008se, lyne_graham-smith_2012}), it is observed that the pulsar rotations are generally stable and show a regular trend in the frequency derivative (spin-down). However, it is a well-established observational fact that the rotational frequencies of certain pulsars are subject to sudden increase, which is often accompanied by a change in the spin-down and an exponential recovery of some fraction of the initial frequency jump~\cite{Lyne2000,Espinoza}. 
These events are generally referred to as \textit{pulsar glitches}~\cite{Link:1992mdl,Link:2000mu,Fuentes:2017bjx,Lyne2000}, which will be referred to in short as glitches in this paper. 
Glitches can produce a non-zero, time-varying quadrupole moment of the NS, and, in turn, lead to GW emission. GWs from the pulsar glitches can be naturally divided into two categories: burst-like GWs, during or shortly after the glitch itself~\cite{Lopez:2022yph,KAGRA:2021tnv,Ho2020}, and continuous GW signals, following the glitch and so-called recovery phase~\cite{vanEysden:2008pd,Keer:2014uva,PhysRevD.95.024022,Prix:2011qv,Yim:2020trr,TCW-prospects-Moragues:2022aaf}.

Out of the many theories proposed so far to explain these events, there are two leading models \cite{Haskell:2015jra, Ruderman:1972aj}, one based on the superfluid pinning model~\cite{AndersonItoh,Alpar} and another related to the crust cracking model~\cite{BAYM1972816}. It has been shown that~\cite{Keer:2014uva} the crust cracking model is unable to describe the largest glitches, like the one for the Vela pulsar~\cite{Alpar96}. Thus, in this paper, we will focus on the superfluid pinning/unpinning model, which is consistent with the observations of two types of glitchers: normal and Vela-like \cite{Ashton:2018qth, Alpar1993ApJ...409..345A, Chau1993ApJ...413L.113C}.

The quantum nature of the superfluids is at the core of the glitch model considered in this paper. According to the model, the NS rotation can be attributed to an array of $\sim 10^{18}$ quantized superfluid vortices \cite{AndersonItoh, Lasky_Review, Andersson:2006nr, Muslimov} that weave the entire NS interior. 
If vortices are strongly attracted or ‘pinned’ to ions in the crust or flux tubes in the core of the star, they cannot move out. 
This pinning restricts their outward movement when the crust spins down~\cite{Khomenko:2018rgj}.
Thus the superfluid core stores a higher angular velocity compared to the crust of the NS. This differential lag builds up between these two components. 
According to the model, a glitch occurs when a few vortices unpin and cause an avalanche of $\sim 10^{7}-10^{15}$ unpinned outward-moving vortices ~\cite{Lasky_Review, WarMel2008:10.1111/j.1365-2966.2008.13662.x}, abruptly transferring the angular momentum to the crust. The vortex avalanche may cause a series of GW bursts \cite{Melatos_2008} during the rise-time of the glitch, and in turn may excite one or more families of the NS global modes (such as f-modes, p-modes, g-modes, and w-modes \cite{Sidery:2009at}), whose GW counterparts are not considered in this work. Finally, due to non-axisymmetric Ekman flow \cite{vanEysden:2008pd}, there may be a continuous periodic signal, close to the NS rotation frequency, that fades during the post-glitch recovery phase.

Searches for continuous waves (CWs) from glitching pulsars \cite{Prix:2011qv, Ashton:2017wui, Ashton:2018qth} are typically performed in three ways: (1) minimally modeling the aftermath of the glitch and searching for ``transient'' CWs~\cite{Prix:2011qv, Keitel:2019zhb, TCW-prospects-Moragues:2022aaf}, (2) ignoring the glitches and analyzing the periods before and after them~\cite{LIGOScientific:2008hfq,LIGOScientific:2019mhs}, and (3) allowing a small mismatch between the electromagnetic (EM) and GW frequency about the time of the glitch~\cite{LIGOScientific:2020lkw}. 
In the most recent observing run of the LIGO-Virgo-KAGRA (LVK) collaboration, CWs due to quadrupolar~\cite{LIGOScientific:2020lkw} and r-mode~\cite{LIGOScientific:2021yby} GW emission from the most active known glitcher, PSR J0537-6910 have been searched for. 
Furthermore, in \cite{cw-narrow-glitchO3-LIGOScientific:2021quq}, a search for transient CWs from six glitching pulsars was also performed, in which a ``window function'' model \cite{Prix:2011qv} was used to define the post-glitch period where GWs are emitted~\cite{Keitel:2019zhb}, resulting in upper limits on the signal strain amplitude as a function of possible post-glitch relaxation duration but not constraining other parameters of the glitches.

GW burst searches are unmodeled analyses that typically hunt short duration \cite{LIGOScientific:2019ryq, LIGOScientific:2022sts, AbadieLIGO2010, Ebersold:2020zah,LIGOScientific:2021nrg} (milliseconds to few seconds) or long duration \cite{long-transient-Thrane:2010ri, long-burst-KAGRA:2021bhs, VanPutten:2016pif,Lai:1994ke, 2011ApJ...736..108P, Piro:2012ax, LIGOScientific:2022jpr} (longer than few seconds) GW transients, whose waveforms are not well-modeled enough and not suited for matched filtering. Pulsar glitches enter in the first category, namely the short-duration bursts. The first direct search for the GW burst counterpart of a pulsar glitch was performed targeting the 2006 Vela pulsar glitch \cite{AbadieLIGO2010}, looking for a signal associated with oscillations of the fundamental quadrupole mode excited by the glitch. Recently, in the third observing run of the LVK collaboration, all-sky searches have been performed for short-duration \cite{AllSky_transient} GW bursts. In all these cases, finding no evidence of GWs, constraints were placed on the individual glitch properties.

Standard GW searches have not been able to detect any CW or burst-like GW signal that can be associated with a single pulsar glitch.
However, there is a third kind of GW signal that can be associated with an ensemble of glitching pulsars: a SGWB. Searching for a SGWB from pulsar glitches can be motivated by the number of known pulsars and observed pulsar glitches in the EM domain. Pulsar catalogs (ATNF\footnote{https://www.atnf.csiro.au/research/pulsar/psrcat/glitchTbl.html} \cite{Manchester:2004bp} and Jodrell Bank\footnote{https://www.jb.man.ac.uk/~pulsar/glitches.html} \cite{Espinoza}) encode the information about the parameters of more than 600 pulsar glitches from a fraction of the known $\sim 3000$ pulsars. This means that, from the discovery of the first pulsar until today, 0.2 glitches per pulsar have been observed. If this proportion is the same when considering the expected $\mathcal{O}(10^8)-\mathcal{O}(10^9)$ \cite{1987ApJ...319..162N, 2007PhyU...50.1123P} neutron stars
in the Milky Way, one can assume that around $2\times \mathcal{O}(10^7)-\mathcal{O}(10^8)$ glitch may have happened in that period of time. 
These numbers suggest that a SGWB could emerge from the superposition of the GW signals from pulsar glitches.

The detection and characterization of this SGWB would allow to provide complementary information to the one from GW and EM searches for individual glitch, since it would give access to properties of glitches and glitching pulsars as a population and does not require individual-glitch observations. In this work, we aim to characterize and constrain the SGWB from the superposition of burst-like GW signals associated with vortex-avalanche during the glitches of Galactic pulsars, assuming the superfluid pinning/unpinning model as glitch source, while being agnostic with respect to the overall number of pulsars in the population. To search for such SGWB, we use cross-correlation methods~\cite{Allen:1997ad, Romano:2016dpx}, which allows us to search for a common signal in multiple data streams simultaneously and disentangle it from instrumental noise. From the results of the search, which does not show any evidence for a SGWB signal, we derive constraints on the average glitch duration $\tau_{\mathrm{av}}$ and the average radial motion of the vortices during the glitches $\Delta r_{\mathrm{av}}$ for the population of the glitching Galactic pulsars. This approach is a novel way to probe the astrophysical properties of pulsar glitches. 

This paper is organized as follows:
in section \ref{sec:SGWB}, we present the model for the SGWB from vortex-avalanches pulsar glitches; in section \ref{sec:methods} we illustrate the search methods; in section \ref{sec:results} we report the results of the analyses. Inferring from these results we also set constraints on the ensemble properties of the pulsar glitches. These are detailed in the same section. We conclude the paper by discussing the implication of these results and the possible extensions in section \ref{sec:conclusions}.

\section{\label{sec:SGWB}Stochastic Gravitational-Wave Background from vortex-avalanches pulsar glitches}
A SGWB arising from the superposition of GW bursts from Galactic pulsar glitches may present characteristic features in space and time domain. It may be expected to follow the angular distribution of the Galactic NSs, which is peaked in the Galactic plane \cite{Lorimer:2012hy}, and hence exhibit anisotropic patterns. In addition, given the large number of glitches and invoking the central limit theorem, the background may be argued to be Gaussian. Nonetheless, given the relatively short duration of the glitches, this may result in it being non-continuous in the time domain. All these aspects are not very known and are worth to be discussed and explored in detail.
However, in this first attempt to search for a SGWB from pulsar glitches using LVK data, we will be working under the simplifying assumption that the SGWB can be described as Gaussian, stationary, and isotropic. We will discuss in section \ref{sec:results} how this may affect the results of the analysis, and we leave the study of the spatial and temporal features of the background for dedicated works in the future.

Under these assumptions, the SGWB can be characterized by measuring and studying the frequency spectrum of $\Omega_{\mathrm{GW}}$, which is the ratio between the GW energy density $\rho_{\mathrm{GW}}$, and the critical energy density needed to have a flat Universe $\rho_{c} \equiv \frac{3 H_0^2 c^2}{8 \pi G}$, with $G$ Newton's gravitational constant, $c$ the speed of light, and $H_0 = 67.9 \, \mathrm{km\, s^{-1} \, Mpc^{-1}}$~\citep{Planck:2015fie} the Hubble parameter today:
\begin{widetext}
\begin{equation}
    \label{eq:omega_iso_general_formula}
    \Omega_{\mathrm{gw}}(f) = \frac{f}{\rho_{c}} \, \dv{\rho_{\mathrm{gw}}(f)}{f},
\end{equation}
where $f$ is the frequency of the GWs.
We consider the above equation in the case of the NS glitches and derive the expression for $\Omega_{\mathrm{gw}}(f)$~\citep{WarszawskiMelatos}.
A detailed calculation is shown in the appendix \ref{sec:signal}.
This leads to the approximation of $\Omega_{\mathrm{gw}}(f)$ as a power-law in frequency for two different regimes of the glitches from unpinning vortices:
\begin{equation}
\label{eq:omega_f_glitches}
        \Omega_{\mathrm{gw}}(f) 
        \approx  \left(\frac{\Theta}{10^2\, \mathrm{s^{-1}}}\right)^2\, \left(\frac{\left\langle1/D^{2}\right\rangle_{\mathrm{NS}}}{1/(6\, \mathrm{kpc})^2}\right)\, \times
        \begin{cases}
            1.09 \times 10^{-27}\, \left(\frac{\left\langle1/\tau^{5}\right\rangle_{\mathrm{NS}}}{1/(10^{-2}\, \mathrm{s})^5}\right)\, \left(\frac{\left\langle\Delta r^2\right\rangle_{\mathrm{NS}}}{(10^{-2}\, \mathrm{m})^2}\right)\,\left(\frac{f}{25\, \mathrm{Hz}}\right)^{5/2} ,
            \qquad &\tilde{\omega} \ll \Delta\tilde{r} \\
            2.74 \times 10^{-17}\, \left(\frac{\left\langle\tau\right\rangle_{\mathrm{NS}}}{10^{-2}\, \mathrm{s}}\right)\,\left(\frac{f}{25\, \mathrm{Hz}}\right)^{17/2}, \qquad &\tilde{\omega}\gg \Delta\tilde{r}
        \end{cases}
\end{equation}
where $\left\langle \dotsm \right\rangle_{\mathrm{NS}}$ denotes the ensemble average over the glitching NS population; $\Theta$ is the total glitch rate of Galactic NSs; $D$ represents the distance of the sources from the observer; $\tau$ is the glitch duration (i.e., the duration of the emitted GW burst during the vortex-avalanche), and $\Delta r$ is the radial displacement of a vortex during a glitch. In the above equation, $\Delta \tilde{r} \equiv \Delta r/R_{s}$ with $R_s = 10^{4}\, \mathrm{m}$ the average NS radius, and $\tilde{\omega} \equiv \omega\tau$ with $\omega$ the NS angular velocity.
\end{widetext}
The two regimes of interest are associated with different conditions on $\tilde{\omega}$ and $\Delta \tilde{r}$. One of them, $\tilde{\omega} \ll \Delta \tilde{r}$ (with $\Delta \tilde{r} \ll 1$), is such that the azimuthal motion of the vortices is negligible compared to their radial one. This happens when the vortex travel time is much shorter compared to the NS rotation period \cite{WarszawskiMelatos}. The second regime corresponds to the condition $\tilde{\omega} \gg \Delta \tilde{r}$ (with $\tilde{\omega}\ll 1$) and reflects a scenario where a larger contribution to GW strain comes from the azimuthal vortex motion with respect to the radial one.

If we consider Galactic sources only and want to get a rough estimate of the intensity of the resulting SGWB, we may adopt the pivot values for the parameters (see \cite{WarszawskiMelatos}) as given in equation \eqref{eq:omega_f_glitches}. In such a way, the resulting SGWB turns out to be smaller than other astrophysical SGWBs~\cite{Regimbau:2011rp, O3_Iso_PhysRevD.104.022004} (such as the one from binary black holes coalescences, expected to be $\Omega_{\mathrm{gw}}(25\,\mathrm{Hz}) \sim 5 \times 10^{-10}$ \cite{O3_pop_paper_LIGOScientific:2021psn}). However, given the large uncertainty in the parameters from the small number of observed glitches from (Galactic) NSs, the amplitude of the background may change drastically, given some observational constraints on the glitch properties. As an example, for the case $\tilde{\omega} \ll \Delta \tilde{r}$, if the (average) glitch duration was $10^{-3}$ or $10^{-4}\, \mathrm{s}$ and the (average) radial displacement $1\, \mathrm{m}$, this would lead to a boost in the SGWB amplitude by a factor of $10^{9}$ and $10^{14}$ respectively, with respect to the result obtained from equation \eqref{eq:omega_f_glitches} for the pivot values.

\section{\label{sec:methods}Search Methods}

\subsection{Cross-correlation statistic and search for SGWB}
We perform the search for a Gaussian, stationary, unpolarized, and isotropic SGWB. To that aim, we analyze the time-series data from the first three observing runs (O1, O2, and O3) of the Advanced LIGO-Hanford (H) and LIGO-Livingston (L) detectors and the Advanced Virgo (V) detector. We first apply time, and frequency domain cuts, identically to what was done in \cite{O3_Iso_PhysRevD.104.022004, O3_iso_data}.
Then, we perform the cross-correlation search, following the procedures outlined below~\cite{O3_Iso_PhysRevD.104.022004}, with a publicly available algorithm implementation~\cite{stochasticM} written in Matlab.

For each ``baseline'', i.e., a detector pair $IJ$ ($I, J =$ H, L, V), we split the time-series output $s_I(t)$ into segments of duration $T$, labeled by $t$, evaluate their Fourier-transforms $\tilde{s}_I(t;\,f)$, and obtain a segment-dependent cross-correlation statistic. Thus, we can define the following ``narrow-band'' cross-correlation estimator at every frequency as~\citep{Romano:2016dpx}
\begin{equation}
    \label{eq:CC_iso}
    \hat{C}_{IJ}(t; f) = \frac{2}{T} \, \frac{\Re[\tilde{s}_{I}^{*}(t; f) \,  \tilde{s}_{J}(t; f)]}{\gamma_{IJ}(f) \, S_{0}(f)}\,,
\end{equation}
where the asterisk $(^*)$ denotes the complex conjugate, $S_{0}(f) = (3H_0^2)/(10 \pi^2 f^3)$, and $\gamma_{IJ}(f)$ is the normalized overlap reduction function~\citep{Allen:1997ad,Christensen_ORF_PhysRevD.46.5250,Flanagan_ORF_PhysRevD.48.2389} that quantifies the reduction in sensitivity due to the geometry of the baseline $IJ$
and its response to the GW signal. 
The normalization of the above defined cross-correlation statistic is chosen in such a way that $\left\langle \hat{C}_{IJ}(t; f) \right\rangle_{\mathrm{time}} = \Omega_{\mathrm{gw}}(f)$ in the absence of correlated noise. 
In the small signal limit, the variance of the above estimator can be expressed as
\begin{equation}
    \label{sigma_CC_iso}
    \sigma^{2}_{IJ}(t; f) \approx \frac{1}{2 T \,\Delta f}\frac{P_I(t; f)P_J(t; f)}{\gamma_{IJ}^2(f) \, S_0^2(t; f)} \,,
\end{equation}
where $P_I(t; f)$ is the one-sided power spectral density in a detector, and $\Delta f$ is the frequency resolution.

Given the broad-band nature of the expected signal, we can obtain the corresponding ``broad-band'' estimator $\hat{C}_{IJ}$ by combining the cross-correlation spectra from different frequencies with appropriate weight factors. This optimal estimator and the associated variance can be expressed as
\begin{align}
    \label{eq:broadband_CC_iso}
    \hat{C}_{IJ} &= \frac{\sum_{k, t} w(f_{\mathrm{k}}) \, \hat{C}_{IJ}(t;\,f_{\mathrm{k}}) \, \sigma^{-2}_{IJ}(t;\, f_{\mathrm{k}})}{\sum_{k} w^2(f_{\mathrm{k}})\sigma^{-2}_{IJ}(t;\, f_{\mathrm{k}})} \,, \\
    \sigma^{-2}_{IJ} &=  \sum_{k, t} w^2(f_{\mathrm{k}}) \, \sigma^{-2}_{IJ}(t; f_{\mathrm{k}}) \,,
\end{align}
where ${f_{\mathrm{k}}}$ is a set of discrete frequencies. The weights $w(f)$ can be derived for a generic $\Omega_{\mathrm{gw}}(f)$ following an optimal filtering approach~\citep{Romano:2016dpx, O3_Iso_PhysRevD.104.022004}
\begin{equation}
    \label{eq:weights_iso}
    w(f) = \frac{\Omega_{\mathrm{gw}}(f)}{\Omega_{\mathrm{gw}}(f_{\mathrm{ref}})} \,,
\end{equation}
where $f_{\mathrm{ref}}$ is an arbitrary reference frequency. In this analysis, we fixed this as $f_{\mathrm{ref}} = 25\; \mathrm{Hz}$ (these choices are in agreement with the one reported in Ref.~\citep{O3_Iso_PhysRevD.104.022004}). 
The optimal estimator and associated variance for a set of individual, independent ($J>I$) baselines can be obtained as follows:
\begin{align}
    \label{eq:CC_iso_final}
    &\hat{C} = \frac{\sum_{IJ} \hat{C}^{IJ} \, \sigma^{-2}_{IJ}}{\sum_{IJ} \sigma^{-2}_{IJ}}\\
    \label{eq:sigma_iso_final}
    &\sigma^{-2} =  \sum_{IJ} \sigma^{-2}_{IJ} \, ,
\end{align}
where results from previous observing runs may be included in the sum as separate baselines. Here, we combine HL-O1, HL-O2, HL-O3, HV-O3, and LV-O3.
Eventually, in the absence of a detection, we can set upper limits on $\Omega_{\mathrm{ref}} \equiv \Omega_{\mathrm{gw}}(f_{\mathrm{ref}})$ through a Bayesian analysis for any model of interest using the estimators presented in equations \eqref{eq:CC_iso_final} and \eqref{eq:sigma_iso_final}. To do that, we employ the likelihood
\begin{equation}
    \label{eq:lklhood_CC}
    p\left( \hat{C}(f_{\mathrm{k}}) | \Omega(f_{\mathrm{k}}) \right) = \frac{1}{\sqrt{2\pi}\sigma(f_{\mathrm{k}})} \,
    \exp\left[-\frac{\left(\hat{C}(f_{\mathrm{k}}) - \Omega(f_{\mathrm{k}})\right)^2}{2\sigma^2(f_{\mathrm{k}})}\right] \,,
\end{equation}
where $\Omega(f_{\mathrm{k}})$ is the model for the SGWB in equation \eqref{eq:omega_f_glitches} and $\hat{C}(f_{\mathrm{k}})$ is assumed to be Gaussian distributed in absence of a signal~\citep{Romano:2016dpx}.
Additionally, we can also use the estimator for $\Omega_{\mathrm{ref}}$
\begin{equation}
    \label{eq:Om_ref_estimator_definition}
    \hat{\Omega}_{\mathrm{ref}}(f_{\mathrm{k}}) \equiv 
    \frac{\hat{C}_{IJ}(f_{\mathrm{k}})}{w(f_{\mathrm{k}})}
\end{equation}
as a starting point to constrain the average glitch duration $\tau_{\mathrm{av}}$ and the average vortex radial displacement $\Delta r_{\mathrm{av}}$ of an ensemble of glitching NSs, which will be discussed next in detail.

\subsection{\label{sec:constraining_parameters}Constraining $\tau_{\mathrm{av}}$ and $\Delta r_{\mathrm{av}}$ from a NS population}
Here, we show how to utilize the results of the cross-correlation search to derive an estimator for an average quantity $q_{\mathrm{av}} \equiv \left\langle q\right\rangle_{\mathrm{NS}}$ of a NS population. 
The method that we present is generic (under some assumptions), and we use it to obtain the estimators for $\tau_{\mathrm{av}}$ and $\Delta r_{\mathrm{av}}$ and constrain them in the two regimes $\tilde{\omega}\ll\Delta\tilde{r}$ and $\tilde{\omega}\gg\Delta\tilde{r}$.

First, we assume that $\Omega_{\mathrm{gw}}(f)$ depends on the quantity of interest $q$ through the ensemble average of its $n$-th power $\left\langle q^{n}\right\rangle_{\mathrm{NS}}$ only. In this way, we can recast equation \eqref{eq:weights_iso} as follows:
\begin{equation}
    \label{eq:om_f_vs_q}
    \Omega_{\mathrm{gw}}(f) = \xi_{q}\, w(f)\, \left\langle q^{n}\right\rangle_{\mathrm{NS}},
\end{equation}
where $\xi_{q} \equiv \xi_{q}({\vec{\pi}}) = \Omega_{\mathrm{ref}}/\left\langle q^{n} \right\rangle_{\mathrm{NS}}$ is a proportionality constant, once the set of parameters characterizing the SGWB ${\vec{\pi}}$ are fixed, while $n \in \mathbb{R}_0$ (if $n=0$, we are estimating $\Omega_{\mathrm{ref}}$, which is already discussed in the previous subsection).
Within this framework, using equation \eqref{eq:CC_iso}, we can rewrite the above equation as
\begin{equation}
\label{eq:q_hat_om_hat}
    \widehat{\left(q^n\right)}_{\mathrm{av}}(f_{\mathrm{k}}) = \frac{1}{\xi_q}\frac{\hat{C}_{IJ}(f_{\mathrm{k}})}{w(f_{\mathrm{k}})} \equiv \frac{\hat{\Omega}_{\mathrm{ref}}(f_{\mathrm{k}})}{\xi_q}\,,
\end{equation}
where $\hat{\Omega}_{\mathrm{ref}}(f_{\mathrm{k}})$ is the narrow-band estimator of $\Omega_{\mathrm{ref}}$, while $ \widehat{\left(q^n\right)}_{\mathrm{av}}(f_{\mathrm{k}})$ is the narrow-band estimator of $\left\langle q^n \right\rangle_{\mathrm{NS}}$ (note that the frequencies $f_{\mathrm{k}}$ in the equation are labels and not a functional dependence).

Now, we want to relate $\widehat{\left(q^n\right)}_{\mathrm{av}}(f_{\mathrm{k}})$ with the narrow-band estimator for $q_{\mathrm{av}}$. This can be achieved by considering the expectation value of $\widehat{\left(q^n\right)}_{\mathrm{av}}(f_{\mathrm{k}})$ and its dependence on $q_{\mathrm{av}}$:
\begin{equation}
    \label{eq:qn_expectation}
    \left\langle  \widehat{\left(q^n\right)}_{\mathrm{av}}(f_{\mathrm{k}}) \right\rangle = \left\langle q^n(f_{\mathrm{k}}) \right\rangle_{\mathrm{NS}} =  (q_{\mathrm{av}})^n(f_{\mathrm{k}}) + \dotsm,
\end{equation}
where $(\dotsm)$ stands for terms involving the intrinsic (central) statistical moments of the statistical distribution of the quantity of interest $q$. As an example for $n = 2$, $\langle  \widehat{(q^{2})}_{\mathrm{av}}(f_{\mathrm{k}}) \rangle = \left\langle q^2(f_{\mathrm{k}}) \right\rangle_{\mathrm{NS}} =  (q_{\mathrm{av}})^{2}(f_{\mathrm{k}}) + \mathrm{Var}(q)(f_{\mathrm{k}})$, and the (unknown) bias is encoded in the (unknown) population variance $\mathrm{Var}(q)$. 
Following this, we define the biased estimator for $q_{\mathrm{av}}$
\begin{equation}
    \label{eq:q_estimator}
    \hat{q}_{\mathrm{av}}(f_{\mathrm{k}}) \equiv 
    \left[\widehat{\left(q^{n}\right)}_{\mathrm{av}}(f_{\mathrm{k}})\right]^{1/n},
\end{equation}
where the bias introduced by the other moments is assumed to be negligible (given $q$ being positive definite and peaked around some reference value). The bias could be accounted for in the case of the observation of SGWB from a population of NS glitches by estimating the higher-order moments of the distribution from individual glitch observations or from theoretical models.
Given the above expression for $\hat{q}_{\mathrm{av}}(f_{\mathrm{k}})$, we can derive its uncertainty $\sigma_{\hat{q}}(f_{\mathrm{k}})$ if we know the likelihood function $p_{q} \left(\hat{q}_{\mathrm{av}}(f_{\mathrm{k}})|q_{\mathrm{av}}(f_{\mathrm{k}})\right)$.
The formula of the likelihood can be obtained in two steps: first, by using equations \eqref{eq:om_f_vs_q}, \eqref{eq:q_hat_om_hat}, and \eqref{eq:q_estimator}, we express $\Omega(f_\mathrm{k})$ and $\hat{C}(f_\mathrm{k})$ as a function of $q_{\mathrm{av}}(f_\mathrm{k})$ and $\hat{q}_{\mathrm{av}}(f_\mathrm{k})$; second, we perform a change of variables in equation \eqref{eq:lklhood_CC}.
\begin{widetext}
In this way, we get the likelihood function for $\hat{q}_{\mathrm{av}}(f_\mathrm{k})$, which is no longer a Gaussian:
\begin{equation}
    \label{eq:lklhood_q}
    p_{q} \left(\hat{q}_{\mathrm{av}}(f_{\mathrm{k}})|q_{\mathrm{av}}(f_{\mathrm{k}})\right) = \sqrt{\frac{2}{\pi}}\, \frac{\left| n \right|\, q_{\mathrm{av}}^{n-1}(f_{\mathrm{k}}) \, \xi_{q}}{ \sigma_{\hat{\Omega}}(f_{\mathrm{k}})} \, \exp\left[-\frac{(\hat{q}^{n}_{\mathrm{av}}(f_{\mathrm{k}}) - q^{n}_{\mathrm{av}}(f_{\mathrm{k}}))^2 \, \xi_q^2}{2\sigma^2_{\hat{\Omega}}(f_{\mathrm{k}})}\right] \,,
\end{equation}
where $\sigma_{\hat{\Omega}}(f_\mathrm{k})$ is the standard deviation corresponding to $\hat{\Omega}_{\mathrm{ref}}(f_\mathrm{k})$. By applying the definition of variance, we then get (we omit frequency labels in the right-hand side of the equation and observe that $n<-2$ or $n>0$)
\begin{equation}
    \label{eq:sigma_q}
    \sigma^2_{\hat{q}}(f_{\mathrm{k}}) 
    =\sqrt{\frac{2}{\pi}} \left(\frac{\sigma_{\hat{\Omega}}}{\xi_{q}}\right)^{2/n} \Gamma\left(\frac{n+2}{n}\right) D_{\left(-\frac{n+2}{n}\right)}\left(z\right) e^{-z^2/4}
    - \left[\sqrt{\frac{2}{\pi}} \left(\frac{\sigma_{\hat{\Omega}}}{\xi_{q}}\right)^{1/n} \Gamma\left(\frac{n+1}{n}\right) D_{\left(-\frac{n+1}{n}\right)}\left(z\right) e^{-z^2/4}\right]^2,
\end{equation}
where $D_{(\nu)}(z)$ is a parabolic cylinder function, and $z \equiv -\hat{q}_{\mathrm{av}}^n\,\xi_{q}/\sigma_{\hat{\Omega}} = -\hat{q}_{\mathrm{av}}^n (f_{\mathrm{k}})\,\xi_{q}/\sigma_{\hat{\Omega}}(f_{\mathrm{k}})$. A derivation of the above equation can be found in appendix \ref{appendixB}.
\end{widetext}
Given $\hat{q}_{\mathrm{av}} (f_\mathrm{k})$ and $\sigma^2_{\hat{q}}(f_{\mathrm{k}})$, assuming $q_{\mathrm{av}}$ to be independent of the frequency, we finally obtain the optimal, broad-band, estimator $\hat{q}_{\mathrm{opt}}$, with the relative uncertainty $\sigma_{q, \, \mathrm{opt}}$ as
\begin{align}
    \label{eq:epsilon_hat_opt}
    &\hat{q}_{\mathrm{opt}} = \frac{\sum_{k} \hat{q}_{\mathrm{av}}(f_{\mathrm{k}})\,\sigma^{-2}_{\hat{q}}(f_{\mathrm{k}})}{\sum_{k} \sigma^{-2}_{\hat{q}}(f_{\mathrm{k}})}\,, \\
    &\sigma_{\hat{q}, \, \mathrm{opt}} = \left[ \sum_{k} \sigma^{-2}_{\hat{q}}(f_{\mathrm{k}}) \right]^{-1/2} \,.
\end{align}

\section{\label{sec:results}Results}
We search for an isotropic SGWB resulting from the superposition of GWs from NS glitches from the Galactic population of pulsars, assuming the model in equation \eqref{eq:omega_f_glitches}, recast in the form $\Omega_{\mathrm{gw}} (f) = \Omega_{\mathrm{ref}}\,(f/f_{\mathrm{ref}})^{\alpha}$. 
We perform the analysis on data from the first three observing runs (O1, O2, and O3) of the Advanced LIGO and Virgo detectors, which are publicly available~\citep{gwosc:LIGOScientific:2019lzm}, and follow the methods described in \cite{O3_Iso_PhysRevD.104.022004}.
Using the results of the search, which do not provide any evidence of a signal, we set constraints on population parameters of the pulsar glitches. The results and the relative implications are presented in the following subsections.
\begin{table*}
    \centering
    \begin{tabular}{ |c|c|c|c| }
        \hline
        $\Omega_{\mathrm{gw}}(f)$ & $\hat{C}^{\mathrm{O1+O2+O3}}$  & $\Omega_{\mathrm{ref}}^{95\%,\, \mathrm{Uniform}}$ & $\Omega_{\mathrm{ref}}^{95\%,\, \mathrm{Log-uniform}}$\\
        \hline
        $\propto f ^{5/2}$ & $(-1.2 \pm 1.5)\times 10^{-9}$ & $2.4\times 10^{-9}$ & $7.5\times 10^{-10}$ \\ 
        \hline
        $\propto f ^{17/2}$ & $(3.8\pm 2.5)\times 10^{-17}$ & $8.3\times 10^{-17}$ & $5.7\times 10^{-17}$ \\ 
        \hline
    \end{tabular}
    \caption{Results of the isotropic search for a SGWB from Galactic-NS glitches using data from the first three LIGO-Virgo-KAGRA observing runs. The first row is relative to the regime $\tilde{\omega} \ll \Delta \tilde{r}$, while the second row is relative to the regime $\tilde{\omega} \gg \Delta \tilde{r}$.
    The four columns are the results from our search, in which the frequency scaling of $\Omega_{\mathrm{gw}}(f)$ (first column), the cross-correlation statistics (second column), and the upper limits on $\Omega_{\mathrm{ref}}$, using a uniform (third column) and log-uniform (fourth column) prior, are reported.}
    \label{tab:iso_results}
\end{table*}

\subsection{\label{sec:results_iso}Search for SGWB}
The search has not found any evidence for a SGWB signal. Hence, we set upper limits on $\Omega_{\mathrm{ref}}$. The results are summarized in table \ref{tab:iso_results}. 
The second column of this table contains the value of the cross-correlation statistic and the associated uncertainty, which have been obtained using equation \eqref{eq:CC_iso_final}. The third and fourth columns show the $95\%$ confidence-level Bayesian upper limits for $\Omega_{\mathrm{ref}}$. 
These upper limits are obtained by marginalizing the likelihood function in equation \eqref{eq:lklhood_CC} over a uniform (third column) and a log-uniform prior (fourth column) on the magnitude of the SGWB. 
The choice of a log-uniform prior may seem the most natural since the $\Omega_{\mathrm{ref}}$ range is expected to span several orders of magnitude.
The log-uniform prior range was chosen to be between $10^{-13} \leq \Omega_{\mathrm{ref}} \leq 10^{-5}$ for $\alpha = 5/2$, and $10^{-20} \leq \Omega_{\mathrm{ref}} \leq 10^{-8}$ for $\alpha = 17/2$. 
The lower bound 
was chosen to be of the same order of magnitude as the expected reach of the next-generation ground-based detectors \cite{O3_Iso_PhysRevD.104.022004, PI_curve_Thrane:2013oya}. The upper bound was chosen in such a way that the upper limits on $\Omega_{\mathrm{ref}}$ did not change noticeably when choosing a broader range, reflecting our lack of information about $\Omega_\mathrm{ref}$ a priori. 
Even though the choice of the uniform prior translates to more conservative upper limits, we have included those results as well, choosing its range to be the same as the log-uniform one. In the case $\alpha = 5/2$, the estimator for $\Omega_{\mathrm{ref}}$ and the upper limits for the uniform prior, are of $\mathcal{O}(10^{-9})$, while the ones from the log-uniform prior are of $\mathcal{O}(10^{-10})$. In the case $\alpha = 17/2$ instead, they are of $\mathcal{O}(10^{-17})$ and are several orders of magnitude smaller than the $\alpha = 5/2$ case and the power-law models considered in Ref.~\cite{O3_Iso_PhysRevD.104.022004}: this is expected for this kind of power-law, given the definition of $\hat{\Omega}_{\mathrm{ref}}$ and equation \eqref{eq:weights_iso}.

\subsection{\label{sec:implications_for_parameters}Implications for $\tau_{\mathrm{av}}$ and $\Delta r_{\mathrm{av}}$}

\subsubsection{$\tilde{\omega} \ll \Delta\tilde{r}$}
\begin{figure}
    \centering
    \includegraphics[width = \columnwidth]{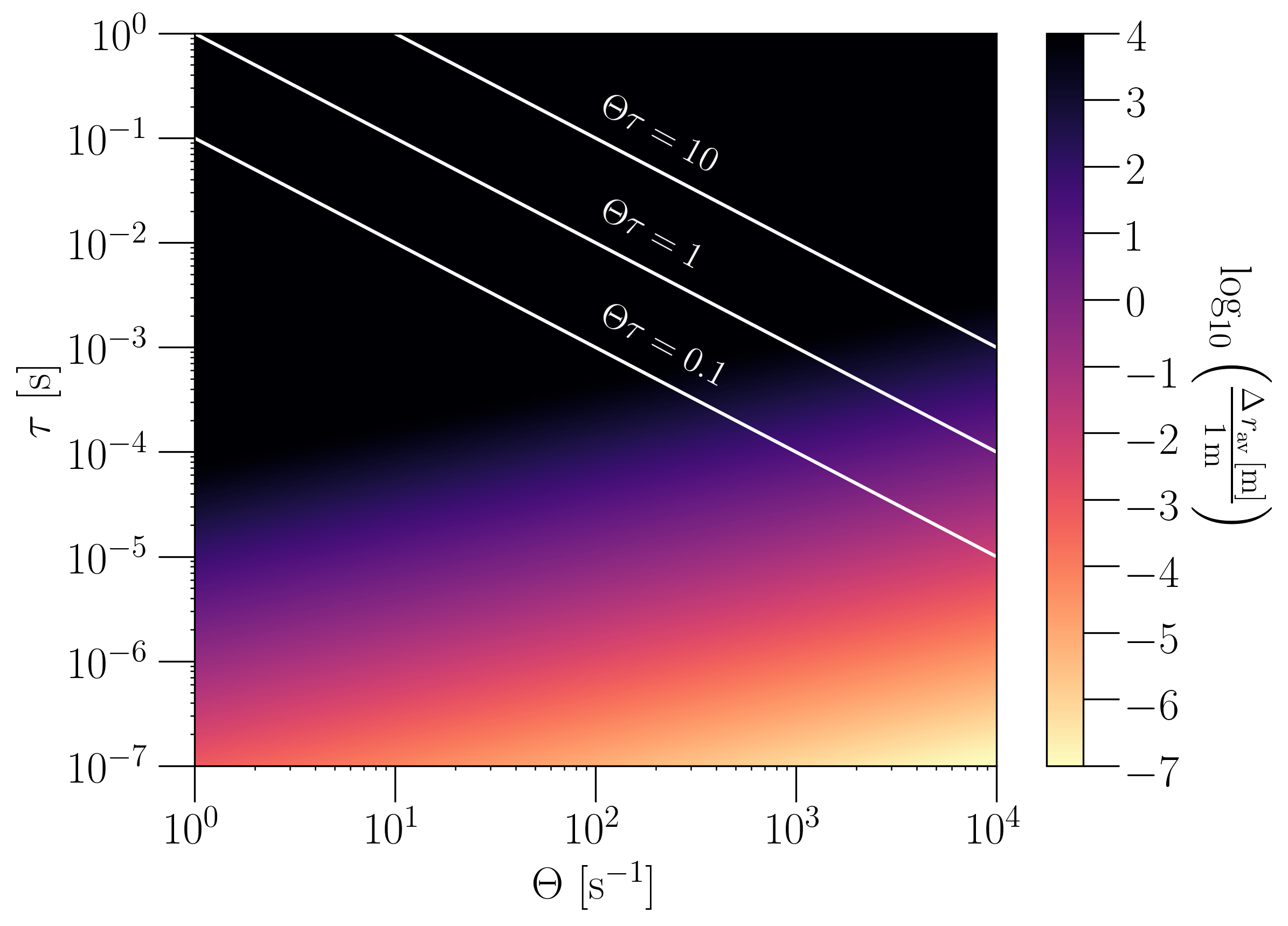}
    \caption{$\tilde{\omega} \ll \Delta\tilde{r}$: 95\% Bayesian upper limits over $\Delta r_{\mathrm{av}}$ in the $\Theta - \tau$ plane, assuming a log-uniform prior on $\Delta r_{\mathrm{av}}$. We choose $\tau\in [10^{-7}, 1]\,\mathrm{s}$ and $\Theta\in [1, 10^{4}]\,\mathrm{s^{-1}}$. The white lines divide the plane into the regions limited by $\Theta\tau = 0.1,\,  1,\,  10$.}
    \label{fig:Delta_r_ULs_alpha2p5}
\end{figure}
\begin{figure}
    \centering
    \includegraphics[width = \columnwidth]{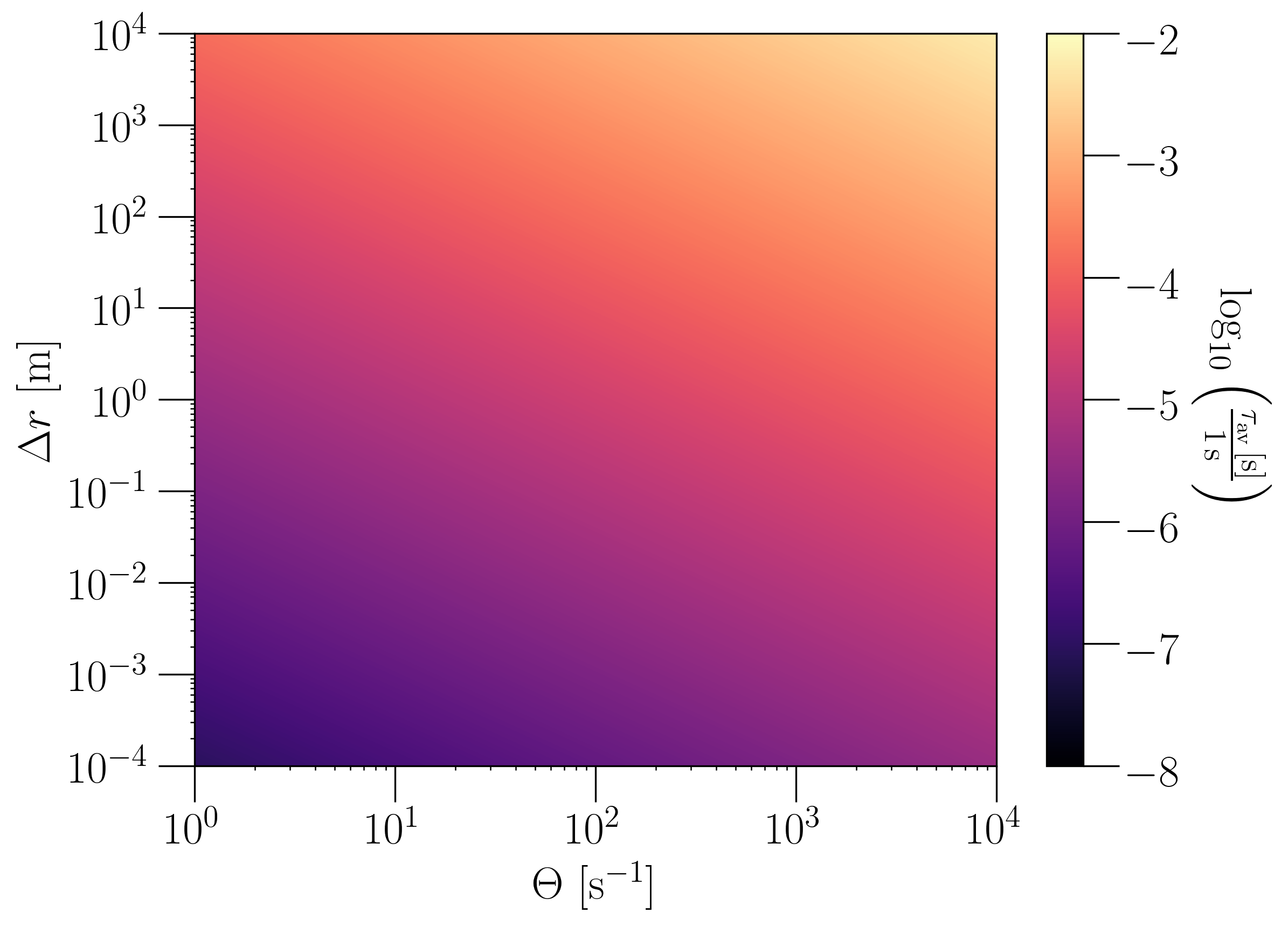}
    \caption{$\tilde{\omega} \ll \Delta\tilde{r}$: 95\% Bayesian lower limits over $\tau_{\mathrm{av}}$ in the $\Theta - \Delta r$ plane, assuming a log-uniform prior on $\tau_{\mathrm{av}}$.We choose $\Delta r \in [10^{-4}, 10^4]\,\mathrm{m}$ and $\Theta\in [1, 10^{4}]\,\mathrm{s^{-1}}$.}
    \label{fig:Tau_ULs_alpha2p5}
\end{figure}
In this regime, by fixing $\left\langle1/D^2\right\rangle^{-1/2}_{\mathrm{NS}}=6\, \mathrm{kpc}$ as the reference value for Galactic pulsars, equation \eqref{eq:omega_f_glitches} depends on three unknown parameters: the total Galactic NS glitching rate $\Theta$, the (effective) glitch duration $\tau \equiv  \left\langle 1/\tau^{5}\right\rangle_{\mathrm{NS}}^{-1/5}$, and the (effective) vortex radial motion $\Delta r \equiv  \left\langle \Delta r^{2}\right\rangle_{\mathrm{NS}}^{1/2}$.
Following the approach described in \ref{sec:constraining_parameters}, we derive Bayesian upper limits for $\Delta r_{\mathrm{av}}$ (keeping $\Theta$ and $\tau$ as free parameters), and $\tau_{\mathrm{av}}$ (keeping $\Theta$ and $\Delta r$ as free parameters). The results are summarized in figures \ref{fig:Delta_r_ULs_alpha2p5} and \ref{fig:Tau_ULs_alpha2p5}.

Figure \ref{fig:Delta_r_ULs_alpha2p5} illustrates the upper limits over $\Delta r_{\mathrm{av}}$ as a function of $\Theta$ and $\tau$, where we choose a log-uniform prior for $\Delta r_{\mathrm{av}}$ with the values ranging in the interval $[10^{-7}, 10^{4}] \, \mathrm{m}$.
The upper bound of the prior is dictated by the fact that $\Delta r_{\mathrm{av}} < R_{S} \simeq 10^{4}\, \mathrm{m}$, while the lower bound is chosen to be small enough that there is no posterior support at the lower end of the prior range.
The constraints over $\Delta r_{\mathrm{av}}$ span the range $(10^{-7} -10^{4})\, \mathrm{m}$, and become more and more stringent the higher the total glitch rate and the lower the average glitch duration are, as expected from equation \eqref{eq:omega_f_glitches}. As a reference, we consider the case where $\Theta = 10^2\, \mathrm{s^{-1}}$ and $\tau = 10^{-2}\, \mathrm{s}$, resulting in $\Delta r_{\mathrm{av}}\leq 9.5 \times 10^{3}\, \mathrm{m}$.
We observe that the upper limits over $\Delta r_{\mathrm{av}}$ in the black region of figure \ref{fig:Delta_r_ULs_alpha2p5} (corresponding to most of the considered parameter space) are not informative, given the information $\Delta r_{\mathrm{av}}\leq 10^{4}\, \mathrm{m}$ is already encoded in the choice of the prior.

Figure \ref{fig:Tau_ULs_alpha2p5} shows the constraints over $\tau_{\mathrm{av}}$, as a function of $\Theta$ and $\Delta r$, where a log-uniform prior with range $[10^{-10}, 10^{2}]\, \mathrm{s}$ is used for $\tau_{\mathrm{av}}$.
The limits over $\tau_{\mathrm{av}}$ are interpreted as lower bounds over the average glitch duration, given $\Omega_{\mathrm{gw}}(f)\propto \left\langle1/\tau^{5}\right\rangle_{\mathrm{NS}}$
as in equation \eqref{eq:omega_f_glitches}.
We note that they cover the range $(10^{-8}-10^{-2})\, \mathrm{s}$, and become more stringent (i.e., the minimal average glitch duration becomes higher) when $\Delta r$ and $\Theta$ increases, as expected again from equation \eqref{eq:omega_f_glitches}. Considering as reference $\Theta = 10^2\, \mathrm{s^{-1}}$ and $\Delta r = 10^{-2}\, \mathrm{m}$, we obtain $\tau_{\mathrm{av}}\geq 3.7 \times 10^{-6}\, \mathrm{s}$. 

It is important to highlight the meaning of regions in which the parameter space is divided in figure \ref{fig:Delta_r_ULs_alpha2p5}. These regions are delimited by the conditions $\Theta\tau = 0.1,\,  1,\,  10$. The quantity $\Theta \tau \equiv \Delta$ is called \textit{duty-cycle} and can be used to infer the statistical properties in the time domain of the SGWB.
Given a collection of events emitting GWs, the duty cycle is defined as the ratio between the average duration of the events and the average time between two successive events.
If $\Delta \gg 1$, the SGWB is continuous and Gaussian, while if $\Delta \ll 1$, the SGWB is shot-noise-like. In the intermediate regime $\Delta \approx 1$, the resulting SGWB is ``popcorn"-like and is no longer Gaussian~\citep{Regimbau:2011rp}. 
The distinction in figure \ref{fig:Delta_r_ULs_alpha2p5} among the regions of the parameter space corresponding to different regimes is relevant to comment on the analysis result, since it assumes the SGWB to be continuous and Gaussian, which means $\Delta \gg 1$. 
In the regions where this condition is not respected, current search techniques may still detect the SGWB, but the overall process requires longer observation periods (see~\cite{Drasco-Flanagan:PhysRevD.67.082003} and section (8.1) in \cite{Romano:2016dpx} for detailed review and discussions). Currently, no searches that allow considering the non-continuous or non-Gaussian nature of the SGWB are available. Still, methods and formalism for such searches have been proposed in the last few years \cite{Smith:2017vfk, Buscicchio:2022raf} and are under development \cite{SSI_search_Lawrence:2023buo, Ballelli:2022bli}.

\subsubsection{$\tilde{\omega} \gg \Delta\tilde{r}$}
\begin{figure}
    \centering
    \includegraphics[width = \columnwidth]{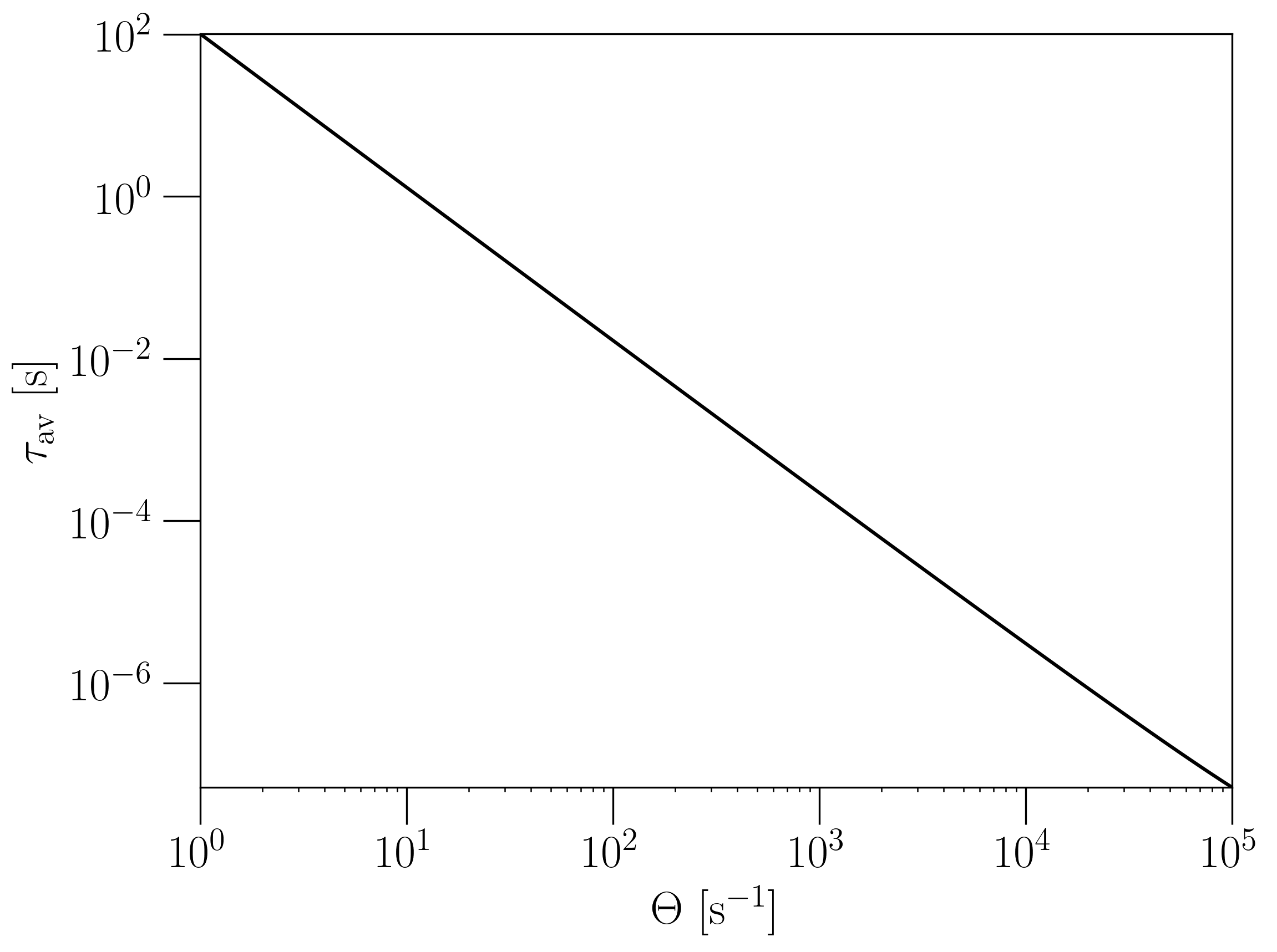}
    \caption{$\tilde{\omega} \gg \Delta\tilde{r}$: 95\% Bayesian upper limits over $\tau_{\mathrm{av}}$ as a function of $\Theta$, assuming a log-uniform prior on $\tau_{\mathrm{av}}$ and choosing $\Theta\in [1, 10^{5}]\,\mathrm{s^{-1}}$.}
    \label{fig:Tau_ULs_alpha6p5}
\end{figure}
In this regime, equation \eqref{eq:omega_f_glitches} does not depend on $\Delta r$. 
Therefore, using the same methods as above, we derive constraints on $\tau_{\mathrm{av}}$ only. We also employ the same priors for $\tau_{\mathrm{av}}$ as in the $\tilde{\omega} \ll \Delta\tilde{r}$ regime.

The results are illustrated in figure \ref{fig:Tau_ULs_alpha6p5}. In contrast to figure \ref{fig:Tau_ULs_alpha2p5}, the limits over $\tau$ in figure \ref{fig:Tau_ULs_alpha6p5} are interpreted as upper limits, given $\Omega_{\mathrm{gw}}(f)\propto \left\langle\tau\right\rangle_{\mathrm{NS}}$.
The constraints span the range $(10^{-7} - 10^{2})\, \mathrm{s}$, and become more stringent with $\Theta$ increasing, in agreement with equation \eqref{eq:omega_f_glitches}. In the reference case with $\Theta = 10^2\, \mathrm{s^{-1}}$, the result is $\tau_{\mathrm{av}} \leq 1.7\times 10^{-2}\,\mathrm{s}$.

\section{\label{sec:conclusions}Discussions and Conclusions}
In this work, we have derived constraints on some ensemble properties of a glitching pulsar population from the results of a cross-correlation-based search for SGWB. 
Throughout this analysis, we have restricted ourselves to the Galactic pulsars, assuming for simplicity the resulting SGWB to be isotropic. We have also considered two glitch regimes in the vortex-avalanche paradigm: $\tilde{\omega} \ll \Delta\tilde{r}$, where the GWs emission is dominated by the radial displacement of the vortices during the avalanche, and $\tilde{\omega} \gg \Delta\tilde{r}$, where the dominant contribution to the GW strain is given by the vortex azimuthal displacement. 
These two regimes give rise to SGWBs that differ in their power-law modeling.
We have not found any evidence in favor of the presence of a SGWB and hence have drawn upper limits on the energy density parameter $\Omega_{\mathrm{gw}}(f)$. These results have been translated into
constraints on the average radial vortex displacement and the average glitch duration as a function of the total glitch rate of Galactic pulsars. 
In the $\tilde{\omega} \ll \Delta\tilde{r}$ regime we have obtained upper limits on $\Delta r_{\mathrm{av}}$ in the range $[10^{-7} \, \mathrm{m}-10^{4}\, \mathrm{m}]$ and lower limits on $\tau_{\mathrm{av}}$ spanning $[10^{-8} \, \mathrm{s}-10^{-2}\, \mathrm{s}]$; while in the $\tilde{\omega} \gg \Delta\tilde{r}$ case we have drawn upper limits on $\tau_{\mathrm{av}}$ in the range $[10^{-7} \, \mathrm{s}-10^{2}\, \mathrm{s}]$.
These results have been obtained using the data from the first three observation runs of Advanced LIGO and Virgo and are the first of their kind.

We have observed that the limits on $\Delta r_{\mathrm{av}}$ become informative (i.e., $\Delta r_{\mathrm{av}} \ll 10^{4}$ m) only in the region of parameter space where the SGWB is non-continuous in the time domain. In addition to that, the lower and upper limits on $\tau_{\mathrm{av}}$ in both regimes confirm that the SGWB is intermittent. On the other hand, the search assumes the SGWB to be continuous in the time domain, which makes it not the most efficient one \cite{Drasco-Flanagan:PhysRevD.67.082003, SSI_search_Lawrence:2023buo} for this specific search for a SGWB from Galactic pulsar glitches. Even though methods optimized for probing such intermittent SGWBs exist \citep{Drasco-Flanagan:PhysRevD.67.082003, Smith:2017vfk, SSI_search_Lawrence:2023buo}, yet no machinery is available as of today.

What we have done in this work can be repeated for other glitch models and may help in constraining their parameters. This approach can be useful to improve our knowledge of (Galactic) pulsar glitch ensemble properties in the absence of direct observations in the GW domain and in spite of the limited number of EM glitching-pulsar observations (several orders of magnitude below the number of pulsars in our Galaxy). 
This indicates that our search, which aims at detecting a SGWB signal, is complementary to the EM observations and the searches looking for GW signals from individual glitches.
First, this search (like other GW searches) is sensitive to quantities that are not accessible (such as the radial motion of the vortices) or are poorly constrained (such as the glitch duration) with EM observations. Second, in contrast to burst-like GWs and CWs searches, the SGWB search has the advantage of instantaneously identifying features ($\Delta r_{\mathrm{av}}$, $\tau_{\mathrm{av}}$) of a known population of glitching pulsars, which would otherwise require decades/centuries to be determined through individual measurements of GW from pulsar glitches. If all the above searches could detect the GWs from pulsar glitches, we would have an ideal platform to implement hierarchical search strategies. 
On one hand, ongoing and future EM-radio missions like UTMOST, MeerKAT, SKA, FAST, and CHIME \cite{bailes_2017, levin_et_al._2017, Jankowski:2018aaq, FAST_2020Innov...100053Q, Lower:2020mjq} would allow the detection of many more glitches and could deliver crucial information about the glitch phase itself, as well as the relaxation phase. On the other hand, information about the glitching pulsars is passed to complementary burst-like and CW-transient GW searches that could, in principle, allow for the multi-messenger astronomy of the glitch phenomenon using data from ground-based GW detectors.
In the case of GW detection, we could combine the parameters determined by these two searches with those of EM observations to improve the measurements and infer the statistical distributions of these quantities, using them as an auxiliary channel to the implications from the SGWB-search.

Finally, we point out some possible ways of extending this work.
As mentioned earlier, throughout this work, we have restricted ourselves to the glitches from Galactic pulsars and assumed the SGWB to be isotropic. If we relax the assumption of isotropy by including the spatial distribution of Galactic NSs, we may perform a targeted search for a SGWB from pulsar glitches that make use of a template-based matched-filtering statistic~\cite{lambda_statistic, Agarwal:2022lvk}.
This approach is expected to provide more insight into pulsar glitch properties. Alternatively, keeping the hypothesis of isotropy, we may consider the resulting SGWB from extra-galactic pulsar populations. In this case, the duty cycle $\Theta \tau \propto D^{3}$, resulting in a SGWB that could be dominated by GWs from extra-galactic pulsar glitches. However, more efforts, including extensive simulations, are required in this case, given our very limited knowledge about these populations. 
Besides these two points, it is also worth mentioning that, recently, there have been several efforts in connecting glitch rates to physical glitch models and modeling the relationships between glitch size and waiting times \cite{Fulgenzi:10.1093/mnras/stx1353, Carlin2020:10.1093/mnras/staa935, Carlin:2021vrr, Haskell:2016ozo, Middleditch:2006ky}, and in linking Galactic pulsar glitch rates to the characteristic age of the pulsars \cite{Millhouse:2022nss}. This additional information, together with the techniques employed in \cite{Millhouse:2022nss}, could be used to estimate $\Theta$ from pulsar catalogs and hence to better model and characterize the SGWB, breaking the degeneracy of $\Theta$ with the other pulsar parameters we have considered in this paper.
These possibilities, together with the effect on the analysis from the inclusion of (future) detectors like KAGRA \citep{KAGRA:2020tym}, LIGO-India \citep{LIGO_India:Saleem_2021}, Einstein Telescope \citep{ET:Punturo:2010zz}, and Cosmic Explorer \citep{CE:Reitze:2019iox} in the detector network, will be considered in future works.

\section*{Acknowledgements}
This material is based upon work supported by NSF's LIGO Laboratory which is a major facility fully funded by the National Science Foundation. This research has used data obtained from the Gravitational Wave Open Science Center, a service of LIGO Laboratory, the LIGO Scientific Collaboration and the Virgo Collaboration. 
We acknowledge the use of Caltech LDAS clusters and the supercomputing facilities of the Universit\'e catholique de Louvain (CISM/UCLouvain) and the Consortium des \'Equipements de Calcul Intensif en F\'ed\'eration Wallonie Bruxelles (C\'ECI), funded by the Fond de la Recherche Scientifique de Belgique (F.R.S.-FNRS) under convention 2.5020.11 and by the Walloon Region. The authors gratefully acknowledge the support of the NSF, STFC, INFN and CNRS for the provision of computational resources. This article has a LIGO document number LIGO-P2200337. 

The authors thank Patrick Meyers for carefully reading the manuscript and providing valuable comments. This work significantly benefited from the interactions with the Stochastic Working Group of the LIGO-Virgo-KAGRA Scientific Collaboration. We also thank L. Warszawski, A. Melatos, T. Regimbau and W. Ho for useful discussions. 

F.D.L. is supported by a FRIA (Fonds pour la formation à la
Recherche dans l'Industrie et dans l'Agriculture) Grant of the Belgian Fund for Research, F.R.S.-FNRS (Fonds de la Recherche Scientifique-FNRS). J. S is supported by a Actions de Recherche Concertées (ARC) grant. A.L.M. is a beneficiary of a Fonds Spéciaux de Recherche (FSR) Incoming Postdoctoral Fellowship. 

We have used {\tt{numpy}} \cite{numpy}, {\tt{scipy}} \cite{scipy}, and {\tt{matplotlib}} \cite{matplotlib} packages to handle data and produce figures of this work.

\bibliography{SGWB_NS_Glitches}

\appendix
\onecolumngrid
\section{\label{sec:signal}Derivation of $\Omega_{\mathrm{gw}}(f)$}
In this appendix, we elaborate the derivation of equation \eqref{eq:om_f_vs_q} for $\Omega_{\mathrm{gw}}(f)$. We start by deriving the expression of the GW strain for a single unpinning vortex. Then we average it over the vortex parameters to get the resulting strain for an individual pulsar glitch. Finally, we apply the definition of $\Omega_{\mathrm{gw}}$, average over the glitch-size distribution of the Galactic NS, and obtain equation \eqref{eq:om_f_vs_q} in the regimes $\tilde{\omega} \ll \Delta \tilde{r}$ and $\tilde{\omega} \gg \Delta \tilde{r}$. The calculation we present below follows \citep{WarszawskiMelatos}, although we correct a few errors which we highlight in our presentation.

\subsection{Single vortex GW signal strain}
The far-field metric perturbation, in the transverse-traceless (TT) gauge, $h_{ij}^{TT}(t)$ can be written as a linear combination of time derivatives of mass multipoles $I^{lm}(t)$ and current multipoles $S^{lm}(t)$. In this work, we neglect the contribution from mass multipoles, assuming the matter distribution inside the pulsar to be incompressible and axisymmetric.
Hence we can rewrite the GW strain as a superposition of the current multipoles only
\begin{equation}
\label{eq:h_current_multipoles}
    h^{TT}_{ij}(t) = \frac{G}{c^4 \, D} \sum^{\infty}_{l = 2} \,
    \sum^{m=l}_{m = -l} T^{\mathrm{B2},\, lm}_{ij} \frac{\partial^{l}\, S^{lm}}{\partial \, t^{l}}(t),
\end{equation}
where $G$ is Newton's gravitational constant, $c$ the speed of light, $D$ the distance from the source to the observer, $T^{\mathrm{B2,\, lm}}_{ij}$ the (``pure spin 2, magnetic type'') tensor spherical harmonics \citep{Thorne:1980ru}, and $t$ the retarded time.
The current multipole moment of order $(l,\, m)$ $S^{lm}(t)$, for a fluid with velocity $\vec{v}$ and density $\rho$ (assumed to be $3\times10^{17}\, \mathrm{kg\, m^{-3}}$), can be expressed as \citep{Thorne:1980ru, Melatos:2009mz}
\begin{equation}
\label{eq:current_multipole_lm}
S^{lm} = \frac{c_{l}}{c^{l-1}} \int_{\mathcal{V}} d^3 x\, Y_{lm}^{*}\,r^{l}\, \vec{x} \cdot \left[\vec{\nabla}  \times (\rho\, \vec{v})\right],   
\end{equation}
where
\begin{equation}
    c_{l} = -\frac{32\,\pi}{\left(2l+1\right)!!}\sqrt{\frac{l+2}{2l\,(l-1)\,(l+1)}},
\end{equation}
and
\begin{equation}
    Y_{lm}(\theta, \varphi) = \sqrt{\frac{(2l+1)\, (l-m)!}{4\pi\,(l+m)!}}\, e^{i\, m\varphi} \, P^{m}_{l}(\cos\theta),
\end{equation}
with $P^{m}_{l}(\cos\theta)$ the associated Legendre function.

The fluid is further assumed to be irrotational and the flow to be purely azimuthal, leading to the following vorticity for vortex singularities \cite{WarszawskiMelatos}:
\begin{equation}
    \vec{\nabla}  \times \vec{v} = \kappa \, \delta^{(2)}\,(\vec{x}_{T} -\vec{x}_{v, \,T}) \hat{z},
\end{equation}
where $\kappa$ is the quantum of circulation, taken to be $\kappa = 10^{-7}\, \mathrm{m^{2}\, s^{-1}}$ \cite{WarszawskiMelatos}, while $\vec{x}_{T}$ and $\vec{x}_{v, \,T}$ are the equatorial coordinate of the NS and the equatorial vortex position.
The leading term to the strain is the current quadrupole $l=2$, whose only non-vanishing terms are $m =\pm 1$ given the above assumptions. 
By switching from Cartesian coordinates $\mathrm{(x, y, z)}$ to the cylindrical ones $(R,\, \phi,\, z)$, it is possible to obtain the following expression
\begin{equation}
\label{eq:current_quadrupole_final}
    S^{21} (t) =\frac{1}c{} \sqrt{\frac{512\pi}{405}} \rho\, \kappa\, e^{-i\, \phi_v(t)}
    R_v(t) \left[R_s^2 - R_v^2 (t)\right]^{3/2},
\end{equation}
where $R_{v}(t) = R_{0} + d(t)$ and $\phi_{v}(t) = \phi_{0} + \omega t$ are the radial and azimuthal positions of the vortex, with $R_0$ and $\phi_{0}$ the positions before unpinning, $R_s$ is the pulsar radius, and $\omega = 2\pi f$ the angular velocity of the pulsar. The radial trajectory $d(t)$ can be modeled as follows
\begin{equation}
    \label{eq:d(t)}
    d(t)=
    \begin{cases}
    0, \qquad &t<t_g\\
    \int_{t_g}^{t_g +t} v(t')\, dt', \qquad &t_g < t < t_g +\tau\\
    \Delta r, \qquad &t>t_g + \tau
    \end{cases}
\end{equation}
where $t_g$ is the time at which the vortex starts moving, $\tau$ is the glitch duration, $\Delta r$ is the (average) radial distance covered by the vortex between unpinning and repinning, and $v(t')$ is the speed profile of the vortex. In the following, we assume a parabolic speed profile $v(t') = 6\, \Delta r\, t'\, (\tau - t')/\tau^3$.
By plugging equation \eqref{eq:current_quadrupole_final} in equation \eqref{eq:h_current_multipoles}, one gets the following expression for the strain from a single vortex ($K_0 \equiv \frac{G}{c^5\, D}\sqrt{\frac{512\pi}{405}}\rho\, \kappa$)
\begin{equation}
\label{h_t_single_vortex}
\begin{split}
    h^{TT}_{ij}(t;\, \omega, \, R_{0}, \, \phi_{0}) &=
    T^{\mathrm{B2,\,21}}_{ij} \, K_{0} \,
    \frac{\partial^2}{\partial t^2}\left[e^{-i \phi_{v}(t)}\,R_v(t) \, \left(R_{S}^2-R_v^2(t)\right)^{3/2}\right],
    \end{split}
\end{equation}
For the calculation in the coming sections, it is useful to express the above equations in terms of the dimensionless variables $\tilde{R}_v(t) \equiv R_v(t)/R_s$, $\tilde{\phi}_v(t) \equiv \phi_v(t)/(2\pi)$, $\tilde{t} \equiv t/\tau$, $\Delta\tilde{r} \equiv \Delta r/R_s$, and $\tilde{f}\equiv f \tau$, hence
\begin{equation}
\label{h_t_single_vortex_tilde}
\begin{split}
    h^{TT}_{ij}(\tilde{t};\, \tilde{\omega}, \, \tilde{R}_{0}, \, \tilde{\phi}_{0}) &=
    T^{\mathrm{B2,\,21}}_{ij} \, \tilde{K}_{0} \, 
    \frac{\partial^2}{\partial \tilde{t}^2}\left[e^{-2\pi i \tilde{\phi}_{v}(\tilde{t})}\,\tilde{R}_v(\tilde {t}) \, \left(1-\tilde{R}_v^2(\tilde{t})\right)^{3/2}\right]
    \equiv T^{\mathrm{B2,\,21}}_{ij} \, \tilde{K}_{0}\, \tilde{h}(\tilde{t}), 
    \end{split}
\end{equation}
where $\tilde{h}(\tilde{t})$ is the second time derivative in the second term, $\tilde{K}_0 \equiv K_0 R_s^4/\tau^2$, $\tilde{R}_0 \in [0,1-\Delta\tilde{r}]$, and $\tilde{\phi}_0 \in [0,1[$, with $\tilde{R}_0 \equiv R_0/R_s$ and $\tilde{\phi}_0 \equiv \phi_0/(2\pi)$.

\subsection{Vortex avalanche signal}
Now that we have the expression for the GW strain from a single vortex, we can derive the one associated with the whole ensemble of unpinning and moving vortices during a glitch. The population properties of the unpinning vortices are related to the glitch geometry, more specifically to the (dimensionless) radial ($\tilde{R_0}$) and azimuthal ($\tilde{\phi}_0$) positions, the final vortex radial displacement ($\Delta\tilde{r}$), and opening angle ($0<\Delta\tilde{\phi}_0<1$) of the vortex avalanche.

Following \citep{WarszawskiMelatos}, the probability distribution functions assumed for the initial positions $\tilde{R}_0$ and $\tilde{\phi}_0$ are
\begin{equation}
    \label{eq:p(R0)}
    p(\tilde{R}_0) = \frac{2\tilde{R}_0}{(1-\Delta \tilde{r})^2}
\end{equation}
and
\begin{equation}
    \label{eq:p(phi0)}
    p(\tilde{\phi}_0)= \frac{H(\tilde{\phi}_0 + \Delta \tilde{\phi}_0/2) - H(\tilde{\phi}_0 - \Delta \tilde{\phi}_0/2)}
    {\Delta \tilde{\phi}_0},
\end{equation}
where $H(\dotsm)$ is the Heaviside step function, with $\tilde{\phi}_{0}$ as bisector of the avalanche.
In this way, we can evaluate the expectation value of the GW strain of a vortex during a glitch as
\begin{equation}
    \label{eq:mu1}
    \left[\mu_1\right]_{ij}(t) = \int_0^{1-\Delta\tilde{r}} \dd\tilde{R}_0\,p(\tilde{R}_0)\,
    \int_{-\Delta \tilde{\phi}_0/2}^{\Delta \tilde{\phi}_0/2}  \dd\tilde{\phi}_0\, p(\tilde{\phi}_0)\,
    h^{TT}_{ij}(t).
\end{equation}
If we take the average over $\tilde{\phi}_{0}$, $\left[\mu_1\right]_{ij}(t)\propto \sin (\Delta\tilde{\phi}_0\,\pi)/(\Delta\tilde{\phi}_0\,\pi)$, which is zero in the case where $\Delta\tilde{\phi}_0=1$, corresponding to the so-called creep-like glitches~\citep{Alpar:1989ApJ...346..823A}, which are not the subject of interest in this paper.
Finally, for a glitch involving the unpinning of $\Delta N_v \gg 1$ vortices, $\left[\mu_1\right]_{ij}(t)$ are drawn from the same Gaussian distribution owing to the central limit theorem, and the GW strain associated with the whole NS glitch $[h_{\mathrm{glitch}}]_{\,ij}(t)$ assumes the form
\begin{equation}
    \label{eq:h_glitch_t}
    \begin{split}
    [h_{\mathrm{glitch}}]_{ij}(t) &= \Delta N_v\, \left[\mu_1\right]_{ij}(t).
    \end{split}
\end{equation}

\subsection{SGWB from vortex-avalanche pulsar glitches}
As discussed in the main text, the superposition of individually undetectable, unresolvable GW signals from NS glitches is expected to generate a SGWB. We recast the general formula for $\Omega_{\mathrm{gw}}(f)$ in equation \eqref{eq:omega_iso_general_formula} in the following simplified form for an astrophysical SGWB \citep{Ferrari:1998jf, Regimbau:2011rp, Phinney:2001di} 
\begin{equation}
    \Omega_{\mathrm{gw}}(f) = \frac{f\, \Theta}{\rho_{c}\, c}
    \frac{\dd[2]{E_{\mathrm{gw}}}}{\dd{f} \dd{A}},
\end{equation}
which can be rewritten as \citep{maggiore2008gravitational}
($\mathcal{F}[\dotsm]$ denotes the Fourier Transform)
\begin{equation}
    \label{eq:omega_strain}
    \Omega_{\mathrm{gw}}(f) = \frac{f\, \Theta\, c^{2}}{32\, \pi\, G\, \rho_{c}}\,
    \left\langle \left|\mathcal{F}[\dot{h}_{ij}] (f)\, \mathcal{F}[\dot{h}^{ij}]^{*} (f)\right| \right\rangle.
\end{equation}
In the present case, $\dot{h}^{ij}$ is the time derivative of $[h_{\mathrm{glitch}}]_{ij}$, and is expressed as
\begin{equation}
\label{eq:dhdt}
    \dot{h}^{ij} (t) = \frac{d[h_{\mathrm{glitch}}]_{\,ij}}{dt}
    \approx T^{\mathrm{B2,\,21}}_{ij} \, K_{0}\,
    \frac{R_s^{4}}{\tau^2}\,
    \frac{\sin (\Delta\tilde{\phi}_0\,\pi)}{\Delta\tilde{\phi}_0\,\pi}\, \Delta N_{v}\, e^{-i\tilde{\omega}\tilde{t}}\, 
    \times \begin{cases}
    \frac{24}{5}\frac{\Delta \tilde{r}}{\tau}, \qquad \tilde{\omega} \ll \Delta\tilde{r}\\
    \frac{i\, \pi \tilde{\omega}^{3}}{16\tau}, \qquad \tilde{\omega} \gg \Delta\tilde{r}
    \end{cases}.
\end{equation}
From equation \eqref{eq:omega_strain}, it appears clear that another ensemble average is necessary over the glitching pulsar population of interest. To that aim, we focus on the $\left\langle \dotsm \right\rangle$ term in equation \eqref{eq:omega_strain}, which is proportional to the squared modulus of equation \eqref{eq:dhdt}, and perform the ensemble average over different parameters.

Given the assumption of isotropy, the averaging over the solid angle involves the tensor spherical harmonics only, and is straightforward, 
given $\int d\Omega_{\hat{n}}\, T^{B2, 21}_{ij}\, (T^{B2,21}_{ij})^{*} = 1$. Similarly, the average over the dimensionless glitch angular opening $\Delta \tilde{\phi}_0$, here assumed to follow a uniform distribution in the $[0,1[$ interval, involves the square of the cardinal sine, translating into an approximate factor equal to 0.451412.

The last quantity to average over in equation \eqref{eq:dhdt} is the number of unpinning vortices $\Delta N_v$. This average procedure can be done more easily in terms of the glitch size $s = \Delta f/f$, which quantifies the variation in the rotation frequency of the pulsar after the glitch, and is related to $\Delta N_v$ approximately as

\begin{equation}
    \label{eq:glitch_size_vs_Nv}
    \frac{\Delta f}{f} \approx \Delta \tilde{r}\frac{\Delta N_v}{N_v}\frac{I_s}{I_c},
\end{equation}
where $I_s/I_c$ is the ratio of the superfluid and crust moments of inertia \citep{Alpar:1989ApJ...346..823A, Link:1992mdl}, and $N_v \approx 4\pi^2 f R_s^2/\kappa$ is the total number of vortices.
We also assume the following power-law distribution for $s$ \citep{harris:1989, porporato:PhysRevE.75.011119, WarszawskiMelatos}, which is supported by the study of multiple glitching pulsars \citep{Melatos_2008} (even though it can be shown not to be universal \citep{Melatos_2008} and there have been several works on its modeling \citep{Fulgenzi:10.1093/mnras/stx1353,Carlin2020:10.1093/mnras/staa935,Carlin:2021vrr, Millhouse:2022nss}):
\begin{equation}
    \label{eq:glitch_size_pdf}
    g(s) = -\frac{1}{2} \left(s_{+}^{-1/2} - s_{-}^{-1/2}\right)^{-1} s^{-3/2},
\end{equation}
where $s_{-}$ and $s_{+}$ are the lower and upper bounds on the glitch size. The upper bound corresponds to when all vortices unpin and can be written as $s_{+} = \Delta \tilde{r}(I_s/I_c)$. At the same time, the lower physical bound $s_{-} \ll s_{+}$ can be estimated by considering the fractional change due to the outward motion of a single vortex, covering a radial distance equal to the inter-vortex separation $\Delta r\approx \sqrt{\kappa/(4 \pi f)}$. From equation \eqref{eq:glitch_size_vs_Nv} one obtains (assuming $I_s / I_c\sim 10^{-2}$)
\begin{equation}
    \label{eq:s_estimate}
    s_{-} = \sqrt{\frac{\kappa}{4\pi f}}\, \frac{k}{4\pi^2 R_{s}^{3}} \frac{I_s}{I_c} \sim 2\times 10^{-30} \left(\frac{f}{100\, \mathrm{Hz}}\right)^{-3/2}.
\end{equation}
Now we proceed in averaging over the glitch size, which results in ($\Delta N_v = s/s_{-}$)
\begin{equation}
    \int_{s_{\mathrm{-}}}^{s_{+}} ds \, g(s)\, \Delta N_v^2
    \approx \frac{1}{3}\frac{s_{+}^{3/2}}{s_{-}^{3/2}} = \frac{N_v^{3/2}}{3} = \frac{1}{3} \left(\frac{f\, R_s^2}{\kappa}\right)^{3/2}.
\end{equation}

Finally, by plugging everything in equation \eqref{eq:omega_strain} and including an additional $\Theta \tau$ to account for the number of simultaneous pulsars glitching during $\tau$ contributing to $\Omega_{\mathrm{gw}}(f)$, we obtain (we drop the $\left\langle...\right\rangle_{\mathrm{NS}}$ in the below equation)
\begin{equation}
\label{eq:omega_f_glitches_appendix}
    \begin{split}
        \Omega_{\mathrm{gw}}(f) 
        \approx 0.451412 \,  \left(\frac{K_0\,R_s^4}{\tau^3}\right)^2\, \left[\frac{1}{3} \left(\frac{f\, R_s^2}{\kappa}\right)^{3/2}\right] (\Theta \tau)\,
         \frac{\Theta\, f\, c^{2}}{32\, \pi\, G\, \rho_{c}}\,\times
        \begin{cases}
             \frac{576}{25}\frac{\Delta \tilde{r}^2}{R_s^2},
            \qquad &\tilde{\omega} \ll \Delta\tilde{r} \\
             \frac{\pi^2 \omega^{6}\, \tau^6}{256}, \qquad &\tilde{\omega}\gg \Delta\tilde{r}
        \end{cases}
    \end{split}
\end{equation}
which is equivalent to equation \eqref{eq:omega_f_glitches} in the main text. 
It is worth noting that the above equation differs from the one reported in Ref. \citep{WarszawskiMelatos} for both the glitch regimes under consideration (also note that the regimes in that equation are swapped compared to how they were originally defined). For the first case, $\tilde{\omega} \ll \Delta\tilde{r}$, the glitch duration dependency of $\Omega_{\mathrm{gw}}(f)$ differs by a factor $\tau^2$. For the other, $\tilde{\omega} \gg \Delta\tilde{r}$, $\Omega_{\mathrm{gw}}(f)\propto f^{17/2}$ compared to $\Omega_{\mathrm{gw}}(f)\propto f^{13/2}$ from Ref. \citep{WarszawskiMelatos}. 
The difference arises from the fact that in Ref. \citep{WarszawskiMelatos} $\Omega_{\mathrm{gw}}(f)\propto \left| \mathcal{F}[(h_{\mathrm{glitch}})_{ij}] \right|^2$ is erroneously used in the very final expression, in contrast to $\Omega_{\mathrm{gw}}(f)\propto \left| \mathcal{F}[(\dot{h}_{\mathrm{glitch}})_{ij}] \right|^2$ here. 

\section{Derivation of Equation \eqref{eq:sigma_q}}
\label{appendixB}
In this appendix, where we omit the frequency labels for compactness reasons, we provide a derivation of the formula for $\sigma_{\hat{q}}$ in equation \eqref{eq:sigma_q}, which is a generalization of equation (3.10) in \cite{Talukder:2014eba}, and equation (B3) in \cite{DeLillo:2022blw}. This can be done by using the definition of variance
$\sigma_{\hat{q}}^{2} \equiv \left\langle q^{2}_{\mathrm{av}}\right\rangle - \left\langle q_{\mathrm{av}}\right\rangle^{2}$, and evaluating the first- and second-order expectation values in the variance definition.
We start by evaluating $\left\langle q_{\mathrm{av}}\right\rangle$:
\begin{align}
    \label{eq:q_exp_value}
    \left\langle q_{\mathrm{av}}\right\rangle &\equiv
    \int_{0}^{\infty} p_{q} \left(\hat{q}_{\mathrm{av}}(f_{\mathrm{k}})|q_{\mathrm{av}}(f_{\mathrm{k}})\right)\, q_{\mathrm{av}} \, \dd{q_{\mathrm{av}}} = 
     \int_{0}^{\infty} \dd{q_{\mathrm{av}}} q_{\mathrm{av}}\, \sqrt{\frac{2}{\pi}}\, \frac{\left| n \right|\, q_{\mathrm{av}}^{n-1} \, \xi_{q}}{ \sigma_{\hat{\Omega}}} \, \exp\left[-\frac{(\hat{q}^{n}_{\mathrm{av}} - q^{n}_{\mathrm{av}})^2 \, \xi_q^2}{2\sigma^2_{\hat{\Omega}}}\right] \nonumber \\
     & = \sqrt{\frac{2}{\pi}} \int_{0}^{\infty} \dd{t}\, t^{1/n}\, \left(\frac{\sigma_{\hat{\Omega}}}{\xi_{q}} \right)^{1/n}\,  \exp\left[-\frac{\left(z+t)^{2}\right)}{2}\right] = \frac{2}{\pi}\, \left(\frac{\sigma_{\hat{\Omega}}}{\xi_{q}} \right)^{1/n}\, \exp\left(- \frac{z^{2}}{4}\right) \int_{0}^{\infty} \dd{t}\, t^{1/n}\, \exp\left(-\frac{t^{2}}{2} -zt - \frac{z^{2}}{4}\right) \nonumber \\
     &\equiv \sqrt{\frac{2}{\pi}}\, \left(\frac{\sigma_{\hat{\Omega}}}{\xi_{q}} \right)^{1/n} \Gamma\left(\frac{n+1}{n}\right)\, \exp\left(- \frac{z^{2}}{4}\right)\, U\left(\frac{n+2}{2n},\, z\right) \nonumber \\
     &= \sqrt{\frac{2}{\pi}}\, \left(\frac{\sigma_{\hat{\Omega}}}{\xi_{q}} \right)^{1/n} \Gamma\left(\frac{n+1}{n}\right)\, \exp\left(- \frac{z^{2}}{4}\right)\, D_{\left(-\frac{n+1}{n}\right)} \left( z\right), \qquad n<-1\; \mathrm{or}\; n>0,
\end{align}
where, in the second line, we have defined $t\equiv q_{\mathrm{av}}^{n}\xi_{q}/\sigma_{\hat{\Omega}}$ and $z\equiv -\hat{q}_{\mathrm{av}}^{n}\xi_{q}/\sigma_{\hat{\Omega}}$, then, in the third line, we have used the following helper function
\begin{equation}
    \label{eq:U_a_z}
    U\left(a, \, z\right) \equiv \frac{\exp{-\frac{z^{2}}{4}}}{\Gamma\left(a+\frac{1}{2}\right)}\, \int_{0}^{\infty} \dd{t} t^{a-1/2}\, \exp \left(-\frac{t^{2}}{2} - zt\right), \qquad \Re{a}>-\frac{1}{2},
\end{equation}
and finally, in the last line, we exploited its relation with the parabolic cylinder function $D_{(\nu)}(z)$ \cite{NIST:DLMF}:
\begin{equation}
    \label{eq:parabolic_U}
    D_{(\nu)}\left(z\right) = U\left(-\nu -\frac{1}{2}, \, z\right).
\end{equation}
On a similar line, one can show that 
\begin{align}
    \label{eq:q_2_exp_value}
    \left\langle q_{\mathrm{av}}^{2}\right\rangle &\equiv
    \int_{0}^{\infty} p_{q} \left(\hat{q}_{\mathrm{av}}(f_{\mathrm{k}})|q_{\mathrm{av}}(f_{\mathrm{k}})\right)\, q_{\mathrm{av}}^{2} \, \dd{q_{\mathrm{av}}} = 
     \int_{0}^{\infty} \dd{q_{\mathrm{av}}} q_{\mathrm{av}}^{2}\, \sqrt{\frac{2}{\pi}}\, \frac{\left| n \right|\, q_{\mathrm{av}}^{n-1} \, \xi_{q}}{ \sigma_{\hat{\Omega}}} \, \exp\left[-\frac{(\hat{q}^{n}_{\mathrm{av}} - q^{n}_{\mathrm{av}})^2 \, \xi_q^2}{2\sigma^2_{\hat{\Omega}}}\right] \nonumber \\
     &= \sqrt{\frac{2}{\pi}}\, \left(\frac{\sigma_{\hat{\Omega}}}{\xi_{q}} \right)^{2/n} \Gamma\left(\frac{n+2}{n}\right)\, \exp\left(- \frac{z^{2}}{4}\right)\, D_{\left(-\frac{n+2}{n}\right)} \left( z\right), \qquad n<-2\; \mathrm{or}\; n>0.
\end{align}
Eventually, but applying the definition of variance, and reinserting the frequency labels, we easily recover equation \eqref{eq:sigma_q}.

\end{document}